  \documentclass[aps,prl,twocolumn,english,superscriptaddress,showpacs,floatfix]{revtex4}

  \usepackage{amsmath,bbm}
  \usepackage{psfig,graphics,graphicx}
  \usepackage{babel}

\begin{document}

  \title{
  How far away is the next earthquake?}

  \author{J\"orn Davidsen}
  \email[]{davidsen@mpipks-dresden.mpg.de}
  \affiliation{Max-Planck-Institut f\"ur Physik Komplexer Systeme,
  N\"othnitzer Strasse 38, 01187 Dresden, Germany}
  \author{Maya Paczuski}
  \email[]{maya@ic.ac.uk}
  \affiliation{Department of Mathematics,
  Imperial College London, London SW7 2AZ, United Kingdom}
  \affiliation{John-von-Neumann Institute for Computing,
  Forschungszentrum J\"ulich, 52425 J\"ulich, Germany}

  \date{\today}

  \begin{abstract}

  Spatial distances between subsequent earthquakes in southern
  California exhibit scale-free statistics, with a critical exponent
  $\delta \approx 0.6$, as well as finite size scaling. The statistics
  are independent of the threshold magnitude as long as the catalog is
  complete, but depend strongly on the temporal ordering of events,
  rather than the geometry of the spatial epicenter distribution.
  Nevertheless, the spatial distance and waiting time between
  subsequent earthquakes are uncorrelated with each other. These
  observations contradict the theory of aftershock zone scaling with
  main shock magnitude.

  \end{abstract}


  \maketitle

  What do we know about earthquakes? Either a great deal or a meager
  amount, depending on the point of view and on the adopted
  definition of an earthquake. If an earthquake is defined to be the
  slip on a fault (or several faults) that produces the observed
  seismic wave field, then we have a good understanding of
  earthquakes \cite{scholz}. In contrast to earthquake kinematics,
  no satisfactory understanding exists of the physical processes in
  the lithosphere that cause slip on faults and are, thus,
  responsible for the \emph{dynamics} of earthquakes
  \cite{mulargia}. This dynamics entangles a vast range of space and
  time scales and manifests itself in a number of generic, empirical
  features of earthquake occurrence including spatio-temporal
  clustering, fault traces and epicenter locations with fractal
  statistics, as well as the Omori and Gutenberg-Richter (GR) laws
  (see Refs.~\cite{turcotte} for a review). The Omori
  law states that the rate of aftershocks after a main shock decays
  hyperbolically in time \cite{omori}, while the GR law
  states that the size (measured in terms of the seismic moment $M$)
  distribution of earthquakes is scale-free \cite{gutenberg}.

  The presence of vastly different scales has  to be
  taken into account in order to interpret measurements correctly. For
  instance, to unambiguously measure the length of many natural
  objects, like the 'fractal' coastline of Norway, one has to specify the
  length of the ruler used \cite{goltz,mandelbrot82}. Here we ask a
  similar question for distances between earthquakes, and find a
  similar result.  That is, in order to determine unambiguously the
  statistics for how far away the next earthquake will be, one has
  to specify the size of the region under consideration.

  We focus on the distribution of spatial distances between the
  epicenters of successive earthquakes in southern California.  In the
  past, different possibilities have been proposed including power-law
  behavior \cite{ito95} and $q$-exponential (cumulative) distributions
  \cite{abe03}. However, none of the previous studies has systematically
  taken into account the physical extent of the region considered, nor
  the threshold magnitude for including events in the analysis. Both of
  these quantities have recently been revealed to be crucial for
  capturing robust, statistical features of waiting times between
  subsequent earthquakes \cite{bak02,corral03,davidsen04}.

  Here, we show that the distribution of spatial distances between
  successive earthquakes, larger than a threshold magnitude $m$,
  occurring within a given region of area $L^2$ exhibits: (1) power law
  behavior with an exponent $\delta \approx 0.6$, (2) finite size
  scaling as a function of $L$ and (3) no dependence on the threshold
  magnitude $m$, as long as the data set is complete for that
  threshold. Our results also provide clear evidence that this behavior
  is not due to a random process bound to the geometrical structure of
  the collection of epicenters, but reflects the complex spatio-temporal organization
  of seismicity. We further argue that the exponent $\delta$ encodes
  information about this complex dynamics, which appears
  unrelated to other known properties of seismicity.  Hence $\delta$
  may be a new, independent exponent characterizing seismicity.

  To analyze the distribution of spatial distances between successive
  earthquakes or ``jumps'', we adopt a method proposed by Bak {\it et
  al.}~\cite{bak02} and take the perspective of statistical physics:
  Neglecting any classification of earthquakes as main shocks,
  foreshocks or aftershocks, analyze seismicity patterns irrespective
  of tectonic features and place all events on the same footing.
  Consider spatial areas and their subdivision into square cells of
  length $L$. For each of these cells, only events above a threshold
  magnitude $m$ are included in the analysis. In this way, we obtain a
  list of the spatial distances $\Delta r_i = \left|{\mathbf
  r}_{i+1}-{\mathbf r}_i\right|$ between successive events with
  \emph{epicenters} ${\mathbf r}_i$ and ${\mathbf r}_{i+1}$ \emph{both
  in the same cell of linear extent} $L$.  Concatenating the lists of
  jumps obtained from each cell, a probability density function of the
  jumps $P_{m,L}(\Delta r)$ can be measured~\cite{note1}.  Since both
  the threshold magnitude and the length scale of the cell are
  arbitrary, we look for robust or universal features of this
  distribution that may appear when these parameters are varied.

  For the SCEDC sub-catalog from southern California we study here,
  the reporting of earthquakes is assumed to be homogeneous from
  January 1984 to December 2000 and complete for events larger than
  magnitude $m_c=2.4$ \cite{wiemer00}. Considering epicenters
  located within the rectangle $(120.5^\circ W, 115.0^\circ
  W)\times(32.5^\circ N, 36.0^\circ N)$ gives $N = 23374$ events
  with magnitude greater than or equal to $m_c$. The GR law for the
  cumulative distribution of earthquakes larger than magnitude $m$
  is $P_>(m) \sim 10^{-bm}$ where the seismic moment $M \propto
  10^{1.5m}$, and $b=0.95$ for this collection of events.

\begin{figure}
   \includegraphics*[width=\columnwidth]{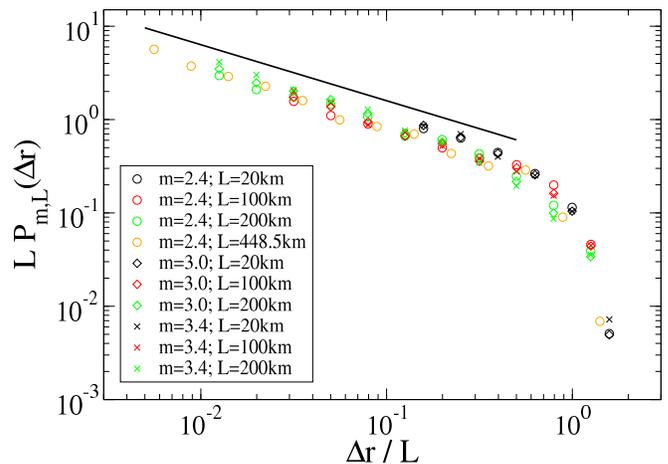}
   \caption{\label{jump}
   (Color online) Rescaled  distribution of jumps, $P_{m,L}(\Delta r)$,
   for different values of $m$ and $L$.
   $L=448.5$km corresponds to the full area where there is just one
   cell. Values of $\Delta r < 2$km have been discarded due to
   uncertainties in estimating the epicenters' locations. The
   solid line with exponent  $\delta = 0.6$ is shown
   as a guide to the eye.
    }
   \end{figure}

  For $m \ge m_c$, we find that the jump distribution is described
  by the finite size scaling (FSS) ansatz
  \begin{equation}\label{scaling}
      P_{m,L}(\Delta r) = \frac{f(\Delta r/L)}{L}\quad ,
  \end{equation}
  where the scaling function $f(x)$ decays as $x^{-\delta}$ with
  $\delta \approx 0.6$ for $x<0.5$
   as shown in Fig.~\ref{jump}. For $x>0.5$, it decays extremely
  rapidly since the finite
  cell size requires that $f(x) = 0$ for $x>\sqrt{2}$.
  These observed results have far reaching implications.

  First, for any $L$ the cutoff sets in at $r \approx L/2$ and, hence,
  scales trivially with $L$. The appearance of FSS precludes the
  existence of any other length scale over the range where FSS holds.  Thus no
  physical length scale exists in the range from 20km to $\approx$
  500km, in contrast to the theory of aftershock zones \cite{kagan02}.
  According to this theory, main shocks generate aftershocks within
  finite aftershock zones, whose extent is comparable to the rupture
  length $l_r = 0.02 \times 10^{0.5\;m}$km of the main event
  \cite{kagan02}.
  This implies that the distance
  between subsequent aftershocks would be limited to the size of the
  largest aftershock zone, which is less than 90km for the catalog
  analyzed here. As in Ref.~\cite{baiesi04}, we
  find no evidence for this length scale.

  Further, as $\delta$ is unambiguously less
  than one, the distribution $P_{m,L}(\Delta r)$ becomes
  non-normalizable for large $L$. Extrapolating our results, the
  finite size of the earth may play an important role in the
  definition of distances between subsequent earthquakes.
  Finally, $P_{m,L}(\Delta r)$ does not significantly depend on $m$
  for $m>m_c$, though the number of included earthquakes varies
  considerably with magnitude.

\begin{figure}
   \includegraphics*[width=\columnwidth]{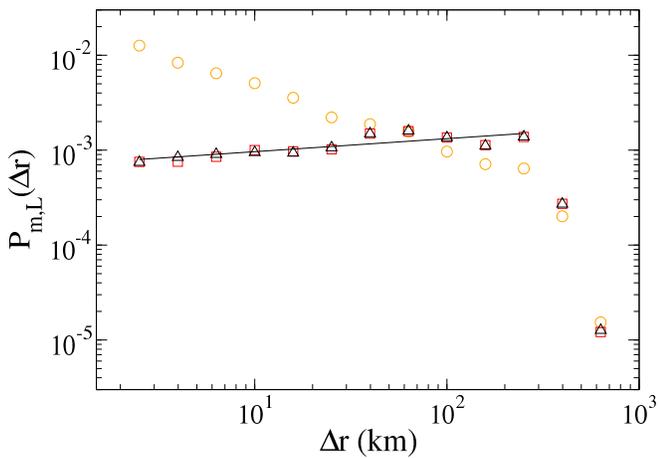}
   \caption{\label{shuff} (Color online) The jump distribution,
   $P_{m,L}(\Delta r)$, for $m=2.4$ and $L=448.5$km.
  (Orange) circles correspond to the distance
   between successive earthquakes $\Delta r_i = \left|{\mathbf
   r}_{i+1}-{\mathbf r}_i\right|$ as in Fig.~\ref{jump}.  (Red) squares
   correspond to $\Delta r_i = \left|{\mathbf r}_{\sigma(i+1)}-{\mathbf
   r}_{\sigma(i)}\right|$ where $\sigma$ is a randomly chosen
   permutation of the integers $0<i<N$, giving a shuffling of the
   catalog as described in the text.  The (black) triangles correspond
   to the distance between \emph{any two} earthquakes in the same cell.
   The solid line is a fit
   to the latter distribution from 2km to 200km with a critical
   exponent of $\delta_{shuf}=-0.14 \pm 0.03$.  }
   \end{figure}

   Although the distribution of jumps reflects a dynamical property of
   seismicity, the particular form of $P_{m,L}(\Delta r)$ could be
   determined by the geometrical structure of the collection of
   epicenters. A simple test can be made by randomly rearranging or
   'shuffling' the temporal sequence of events, while holding the
   magnitude of the events and their epicenters fixed. The
   distribution of distances between subsequent events in the shuffled
   catalog is very different from the original one, as shown in
   Fig.~\ref{shuff}.  For the shuffled catalog, the distribution of
   jumps does not decay with an exponent $\delta$, but it actually
   increases as a power law with exponent $\delta_{shuf} \simeq
   -0.14$. This can be understood from the fact that the
   randomly rearranged ordering gives an estimate of the
   distribution of distances between any two earthquakes
   within the same cell, which is also shown in Fig.~\ref{shuff}. The
   critical exponent of the cumulative distribution is, by definition,
   a measure of the correlation dimension $D_2$ of the epicenter
   distribution \cite{grassberger83}. Our findings imply a fractal dimension
   $D_2=1-\delta_{shuf} \approx 1.14$, which agrees, within
   statistical error, with the value obtained in \cite{davidsen04}.

   The comparison between the distribution of jumps and the
   distribution of distances between epicenter pairs clearly proves
   the dynamical origin of a non-trivial $\delta$. In particular,
   this exponent is
   not simply related to the correlation dimension. Yet, $D_2$ is
   also independent of the threshold magnitude of the
   earthquakes considered and the size of the region studied
   \cite{kagan80}. The latter fact and the appearance of FSS in all
   distributions of
   Fig.~\ref{shuff}, are inconsistent with the existence
   of any preferred length scale, other than the cell size chosen by
   the observer.

  The "propagation" of seismic activity is not only described by
  spatial distances $\Delta r_i$ but also by the waiting time
  $\Delta t_i$ between successive earthquakes $i$ and $i+1$.
  Although the statistics of the waiting times has been studied
  recently \cite{bak02,corral03,davidsen04}, the propagation itself
  has not been analyzed. Here we consider the velocities,
  $v_i=\Delta r_i / \Delta t_i$, between subsequent events in each
  cell for events with magnitude above a threshold $m$ and combine
  them, as before, into a probability density function $P_{m,L}(v)$.
  (Note that $P_{m,L}(v)$ is not directly related to the
  controversial and debated  subject of aftershock diffusion, which
  refers to the expansion or migration of aftershock zones with
  time.  See Ref.~\cite{helmstetter02} for a review.)

\begin{figure}
   \includegraphics*[width=\columnwidth]{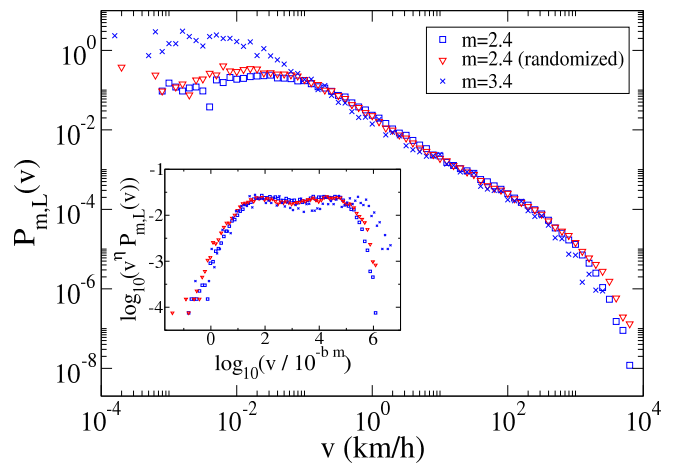}
   \caption{\label{vel} (Color online) The velocity distribution,
   $P_{m,L}(v)$, for $L=100$km and different values of $m$: (Blue)
   crosses and boxes correspond to the original data with $v_i=\Delta
   r_i / \Delta t_i$. (Red) triangles correspond to a randomization
   $v_i=\Delta r_{\sigma(i)} / \Delta t_i$, where $\sigma$ is a
   randomly chosen
   permutation of the integers $0<i<N$. Note that $\Delta r_i >2$km and
   $\Delta t_i>60$s for all $i$ to avoid any bias due to the
   uncertainties in the estimates.  Inset: Rescaled distributions, using
   the GR exponent $b=0.95$ and $\eta=1.0$. }
   \end{figure}

  Figure~\ref{vel} shows that $P_{m,L}(v) \sim v^{-\eta}$ with $\eta
  \approx 1.0$ for intermediate $v$. The cutoff at large $v$ is
  determined solely by $L$, due to the fixed temporal resolution of
  $\Delta t>60$ sec implying $v_{max} = \sqrt{2} \cdot 100$km $\cdot
  60$h$^{-1} \approx 8485$km/h. The cutoff at small $v$ depends on $m$.
  The inset of Fig.~\ref{vel} shows that its position scales as
  $10^{-b\;m}$.
  If we ignore, for the moment, the variation of distances
  $\Delta r$, then $P_{m,L}(v) \rightarrow P_{m,L}(1/\Delta t)$.  The
  latter distribution is obtained from the distribution of waiting times
  $P_{m,L}(\Delta t)$ in \cite{bak02,corral03,davidsen04}.  Since
  $P_{m,L}(1/\Delta t) \, d(1/\Delta t) = P_{m,L}(\Delta t) \, d\Delta
  t$, in this approximation the exponent $\eta = 2 - \alpha$, where
  $\alpha \approx 1$ is the exponent characterizing the distribution of
  waiting times for intermediate arguments.  Furthermore, the cutoff at
  small $v$ would be controlled by the cutoff at large $t$ in the
  waiting time distribution where the behavior crosses over from a
  power-law regime with exponent $\alpha \approx 1$ to a faster decay
  at $t_{cutoff} \sim 10^{b\;m}$ \cite{bak02,corral03,davidsen04}.

  {\it If the statistics of the waiting times and
  the jumps are independent, then $P_{m,L}(v)$ will only reflect the
  statistics of the waiting times over a range of velocities.} This is
  due to the fact that the distribution of waiting times is much broader
  (approximately seven orders of magnitude) than the distribution of
  spatial distance (approximately three orders of magnitude) for our
  data.
  Indeed, the particular combination of
  spatial and temporal distances between successive earthquakes is
  largely random.  As shown in Fig. 3, the
  estimate of $P_{m,L}(v)$ does not change significantly if $v_i$ is
  given by $v_i=\Delta r_{\sigma(i)} / \Delta t_i$ where $\sigma$ is
  a random permutation of the integers $0<i<N$. This clearly
  proves that waiting times and jumps are independent of each other.
  Additionally, using  the same values of $m$ and $L$ as  in
  Fig.~\ref{vel}, the spatial and temporal distances between
  successive earthquakes are almost uncorrelated as measured by the
  cross correlation
  $$r \equiv \frac{\langle (\Delta r_i - \langle
  \Delta r_i \rangle_i) \cdot (\Delta t_i - \langle \Delta t_i
  \rangle_i) \rangle_i}{\sqrt{\langle (\Delta r_i - \langle \Delta
  r_i \rangle_i)^2 \rangle_i} \sqrt{\langle (\Delta t_i - \langle
  \Delta t_i \rangle_i)^2 \rangle_i}} \approx 0.07 \quad , $$ where
  $\langle \dots \rangle_i$ denotes an average over events $i$.

  Considering the observation that the critical exponent $\delta$
  depends on the temporal order of events and is not determined
  solely by the geometry of the set of epicenter locations, together
  with the observation that the waiting times and jumps are largely
  uncorrelated and apparently independent of each other suggests
  that the exponent $\delta$ may
  be a new, independent exponent characterizing seismicity.

  The lack of correlation between waiting times and jumps could be
  interpreted as an indication that the aftershock decay rate at all
  distances is the same, and could therefore have implications both
  for models of aftershock diffusion and for the rate and state/static
  stress friction model of aftershock triggering \cite{helmstetter02}.
  However, our analysis is not based on the distinction between main
  shocks or aftershocks. This distinction is relative
  ~\cite{baiesi04,kagan_quote}.  In fact, there is no
  unique operational way to distinguish between aftershocks and main
  shocks \cite{bak02} and they are not caused by different relaxation
  mechanisms
  \cite{houghs97,helmstetter03}.
  Besides, such a classification may not always be the most
  useful way to describe the dynamics of seismicity.  One could
  nevertheless consider the possibility that the scaling region in
  Fig.~\ref{jump} can be attributed entirely to aftershocks. This
  would require that aftershock sequences dominate the statistics
  during the period and magnitude range considered.
  According to the traditional classification and using
  aftershock zone scaling with main magnitude, the maximum
  distance between aftershocks would be
  determined by the largest events, namely the $m=7.3$ Landers
  earthquake and the $m=7.1$ Hector
  mine earthquake. Thus, this distance should be less than 90km.
  However, we find no break or change in
  scaling behavior for larger distances extending all the way up to
  the size of the region consider, of the order of 500km.

  Our results are strikingly different from earlier results by Ito
  \cite{ito95} and Abe and Suzuki \cite{abe03} for California, who
  examined similar catalogues over a similar time span but did not
  take into account the length scale of observation.  The latter
  authors also included earthquakes with magnitude as low as 0.0,
  and (as in \cite{ito95}) found a very different jump distribution,
  implying that their analysis suffered from the incompleteness of
  the earthquake catalog at small magnitudes. Although the
  distribution of jumps is independent of the threshold magnitude as
  long as the catalog is complete, the marked difference between our
  results and previous ones shows that the completeness of
  earthquake catalogs is crucial for obtaining robust and accurate
  results.

  As pointed out by Corral \cite{corral03,corral04}, the (long-term)
  measurement of the distribution of waiting times between
  subsequent earthquakes --- as in Ref.~\cite{bak02,davidsen04} ---
  involves an average over regions and time spans with widely
  different rates of seismic activity. If, instead, statistics are
  measured in regions and time spans with a stationary rate of
  earthquakes, a different distribution is obtained. A similar
  situation occurs for our measurement. The universal law we find
  for spatial distances between subsequent events holds for data
  sets where the rate of earthquakes is heterogeneous, i.e., for
  rather long time spans. Analysis at short time scales in the
  stationary regime, as described in Ref.~\cite{corral03} will be
  explored in a future work.

  To summarize, we have shown that the distribution of spatial
  distances between successive earthquakes in southern California
  exhibits finite size scaling with a non-trivial power law exponent
  $\delta \approx 0.6$.  Thus, in
  order to specify the statistics of distances between
  earthquakes, one has to define the length scale of observation,
  since no intrinsic length scale exists. This
  observation is in sharp contrast to the theory of aftershock zone
  scaling
  with main shock magnitude
  \cite{kagan02}.  The exponent $\delta$ has a dynamical origin,
  but the distances and waiting times between subsequent events
  are found to be independent of each other.
  This implies that the complex dynamics
  of seismicity has a self-similar hierarchical structure in space
  and time, consistent with the hypothesis that it is a
  self-organized critical phenomenon \cite{bak89}.  Our findings can
  be used as benchmark tests for models of seismicity.

  We thank the Southern California Earthquake Data Center for providing
  the data, and M. Baiesi for helpful conversations. JD would like to
  thank Imperial College London for its hospitality.


\begin{thebibliography}{21}
\bibitem{note1}
  Neither $P_{m,L}(\Delta r)$ nor the other distance distributions
  studied here are  probability densities  per unit area. To obtain the
  latter, one has to divide the respective former distribution by $2
  \pi \Delta r$ and normalize it appropriately.
\end{thebibliography}
  \end{document}